\documentclass[11pt,twoside,a4paper]{article}
\usepackage[affil-it]{authblk}
\usepackage{amssymb }
\usepackage{amsmath}
\usepackage{tensor}
\usepackage{color}
\usepackage{tikz}
\usetikzlibrary{matrix}
\usepackage[a4paper]{geometry}
\usepackage{polski}
\usepackage[cp1250]{inputenc}
\usepackage[polish,english]{babel}
\usepackage[toc,page]{appendix}
\usepackage{pgfgantt}
\usepackage{enumitem}
\usepackage{marvosym}
\usepackage{scrextend}
\usepackage{titling}

\input xy
\xyoption{all} \tolerance=500

\newdir{ (}{{}*!/-5pt/@^{(}}
\newdir{|>}{!/4,5pt/@{|}*:(1,-.2)@^{>}*:(1,+.2)@_{>}}

\def\ppt{\frac{\partial}{\partial t}}

\def\wdt{\widetilde }

\def\mb{\mathbb}

\newtheorem{theorem}{Theorem}
\newtheorem{definition}{Definition}

\usepackage{xcolor,colortbl}

\usepackage{forloop}
\newcounter{loopcntr}

\usepackage{hyperref}

\title{Time-dependent Hamiltonian mechanics on a locally conformal symplectic manifold}
\date{}

\begin{document}

\author{Orlando Ragnisco$^{\dagger}$, Cristina
Sard\'on$^{*}$,  Marcin Zaj\k{a}c$^{**}$}
\date{}
\maketitle

\vskip -0.5cm

\centerline{Department of Mathematics and Physics$^{\dagger}$,}
\centerline{Universita degli studi Roma Tre,}
\centerline{Largo S. Leonardo Murialdo, 1, 00146 , Rome, Italy.}
\centerline{ragnisco@fis.uniroma3.it}
\vskip 0.5cm

\centerline{Department of Applied Mathematics$^{*}$,}
\centerline{Universidad Polit\'ecnica de Madrid.}
\centerline{C/ Jos\'e Guti\'errez Abascal, 2, 28006, Madrid. Spain.}
\centerline{mariacristina.sardon@upm.es}
\vskip 0.5cm

\centerline{Department of Mathematical Methods in Physics$^{**}$,}
\centerline{Faculty of Physics. University of Warsaw,}
\centerline{ul. Pasteura 5, 02-093 Warsaw, Poland.}
\centerline{marcin.zajac@fuw.edu.pl}

\vskip 1.0cm

\begin{abstract}
\noindent
In this paper we aim at presenting a concise but also comprehensive study of time-dependent ($t$-dependent) Hamiltonian dynamics on a locally conformal symplectic (lcs) manifold. We present a generalized geometric theory of canonical transformations and formulate a time-dependent geometric Hamilton-Jacobi theory on lcs manifolds. In contrast to previous papers concerning locally conformal symplectic manifolds, here the introduction of the time dependency brings out interesting geometric properties, as it is the introduction of contact geometry in locally symplectic patches. To conclude, we show examples of the applications of our formalism, in particular, we present systems of differential equations with time-dependent parameters, which admit different physical interpretations as we shall point out. 

\end{abstract}

\tableofcontents

\section{Introduction}




Symplectic geometry has been widely developed since its beginnings in 1940's, the literature on this topic is very vast, and there exist very complete anthologies on it \cite{Arnold,GotayIsenberg,McDuff}. On the contrary, the studies on locally conformal symplectic structures (lcs henceforth) are not so extended. Except for the works of Libermann in 1955 \cite{Liberm} and Jean Lefebvre from 1966--1969 \cite{Lefebvre1,Lefebvre2}, the
subject remained untouched until Vaisman revived the topic in the 70's \cite{vaisman,vaisman2}. One can notice though that recently, the topic of lcs manifolds has become increasingly popular. For an extended description of the revival of lcs geometry and a comprehensive discussion of its relations to other topics, we refer the reader to \cite{Banzoni}.


Let us refresh the memory of the reader with a brief recalling of the setting of lcs geometry more explicitly. The advent of lcs was rooted in Hwa-Chung Lee's works in 1941 \cite{Hwa}, which reconsider the general setting of a manifold endowed with a non-degenerate two-form $\omega$. First, he discussed the symplectic case, and then the problem of a pair of two-forms $\omega_1$ and $\omega_2$, which are conformal to one another. 
Later in 1985, Vaisman \cite{vaisman} defined a lcs manifold
as an even dimensional manifold endowed with a non-degenerate two-form such that for every point $p\in M$ there is an
open neighborhood $U$ in which the following equation is satisfied:
\begin{equation} \label{cond}
d\left(e^{-\sigma}\omega|_{U}\right)=0,
\end{equation}
where $\sigma:U\rightarrow \mathbb{R}$ is a smooth function. If this condition holds for every point of $M$, then $(M,\omega)$ is called a globally
conformal symplectic manifold (gcs manifold). If \eqref{cond} holds for a constant function $\sigma$, then $(M,\omega)$ is a symplectic
manifold. The work of Lee proposes an equivalent definition with the aid of
a compatible one-form, named the Lee form  \cite{Hwa}, that we will introduce  in subsequent sections. 
It is important to notice that at a local scale, a symplectic manifold can not be
distinguished from a lcs manifold. Thus, not only all symplectic manifolds locally look alike, but there exist
manifolds which locally behave like symplectic manifolds, but they fail to do so globally \cite{Banzoni}.


Locally conformal symplectic structures of  first kind are strictly related to contact structures. Recall that a (co-orientable) contact manifold is the pair $(P,\alpha)$, where $P$ is an odd dimensional manifold, such that $\dim P\geq 3$, and $\alpha$ is a one-form such that $\alpha\wedge(d\alpha)^n\neq 0$ at every point \cite{Geiges,LeonValc}. Frobenius' integrability
theorem shows that the distribution $\xi =\ker\alpha$ is then maximally non-integrable. Let $(P,\alpha)$ be a contact manifold and consider a strict contactomorphism, that is, a diffeomorphism $\phi:P\to P$ satisfying $\phi^*\alpha =\alpha$. Then, as observed for instance by Banyaga in \cite{Banyag}, the mapping torus $P_\phi$ admits a locally conformal symplectic structure of the first kind. A similar result, in which the given lcs structure is preserved, is proved in \cite{BazzonMarrer}.

\noindent
It is interesting to notice that contact and locally conformal symplectic structures converge towards Jacobi structures too. Indeed, a transitive Jacobi manifold is a contact manifold if it is odd dimensional, and a lcs manifold if it is even dimensional \cite{Lichnerow}. Therefore, locally conformal symplectic structures may be seen as the even-dimensional counterpart of contact geometry. 

Concerning the physical applications of lcs structures, it has been proven that they conform a very successful geometric model for certain physical problems. In \cite{MacPrzybTsig} Maciejewski, Przybylska and Tsiganov considered conformal Hamiltonian vector
fields in the theory of bi-Hamiltonian systems in order to reproduce examples of completely integrable systems. In \cite{Marle}, Marle used conformally Hamiltonian vector fields to study certain diffeomorphism between the phase space of the Kepler problem and an open subset of
the cotangent bundle of $S^3$ (resp. of a 2-sheeted hyperboloid, according to the energy of the motion).
In \cite{WoLi98}, Wojtkowski and Liverani apply lcs structures in order to model concrete physical situations such as Gaussian isokinetic dynamics with collisions, and Nosé-Hoover dynamics. More precisely, the authors show that such systems are tractable by means of conformal Hamiltonian dynamics, and explain how one deduces properties of the symmetric Lyapunov spectrum. Another interesting role of lcs geometry is its appearance in dissipative systems. It is possible to present a KAM theory for them and find a solution with a fixed $n$-dimensional (Diophantine) frequency by adjusting the parameters when both the number of degrees of freedom and the number of parameters are $n$ \cite{CallCell}.


Another important structure having a close connection with lcs structures are contact pairs, which we shall discuss in the last section of the paper. Contact pairs were introduced in~\cite{Bande}, although they had been previously discussed in \cite{Abe} with the name of bicontact structures in the context of Hermitian geometry \cite{BLY}.  
In short, a contact pair on a manifold is a pair of one-forms $\alpha_1$ and $\alpha_2$ of constant and complementary
classes, for which $\alpha_1$ induces a contact form on the leaves of the characteristic foliation of $\alpha_2$, and
viceversa. In \cite{Bande,Bande2,Bande3}, it is shown how a locally conformal symplectic structure can be constructed out of a contact pair. This fact is quite remarkable to find physical interpretations. Indeed, our modus operandi consists of identifying Lie algebras that admit a contact pair and construct a Lie system \cite{BBHLS, BCHLS, BHLS, CLS, LS} from such Lie algebras. Lie systems are time-dependent systems of differential equations that admit finite-dimensional Lie algebras, and they usually admit a plethora of applications in biology, economics, cosmology, control theory and engineering. Furthermore, some of these systems admit a Hamiltonian. In this line, one is able to construct time-dependent Hamiltonians that are locally conformal symplectic. The last section of this manuscript is dedicated to applications that will illustrate the mentioned constructions.


The aim of this paper is to present a comprehensive formulation of a time-dependent ($t$-dependent) Hamiltonian dynamics on a locally conformal symplectic manifold. We extend the time-independent formalism explained in \cite{vaisman} to the time-dependent case. Subsequently, we extend the theory of canonical transformations and the Hamilton-Jacobi theory (HJ theory) for time-dependent Hamiltonian systems through lcs geometry. All these new results open a path to study physical systems as the ones above mentioned when one considers their time-dependent version, altogether with the application of advanced methods such as canonical transformations and HJ theory. We approach our study from a local and a global point of view and see the relation between both approaches.

The extension of the notion of time-dependent canonical transformations to lcs structures is naturally led by the relation between the symplectic and the lcs case. Whatsmore, the notion of a canonical transformation is strongly related to the HJ theory, which is another problem of our paper. The HJ theory is a powerful tool in mechanics which has many applications both in classical and quantum physics. The Hamilton-Jacobi equation is one of the most elegant approaches to Lagrangian systems such as geometric optics and classical mechanics, establishing the duality between trajectories and waves and paving the way naturally for quantum mechanics. It is particularly useful in identifying conserved quantities for mechanical systems, which may be possible even when the mechanical problem itself cannot be solved completely. In quantum mechanics it provides a relation between classical and quantum mechanics by means of the correspondence principle, although we shall restrict ourselves to the classical approach here. Indeed, we will take the classical geometric detour, in which the solution to the Hamilton--Jacobi equation is interpreted as a section of the cotangent bundle. 

Let us explain this geometrically through a primordial observation on a symplectic phase space (let us take a symplectic manifold to enunciate the philosophy of the geometric Hamilton--Jacobi theory with the simplest example).  If a Hamiltonian
vector field $X_{H}:T^{*}Q\rightarrow TT^{*}Q$ can be projected into a vector field $X_H^{dW}:Q\rightarrow TQ$ on a lower
dimensional manifold by means of a 1-form $dW$, then the integral curves of the projected
vector field $X_{H}^{dW}$ can be transformed into integral curves of $X_{H}$ provided that $W$ is a solution of the Hamilton--Jacobi equation. Let us define the projected vector field as
\begin{equation}
 X_H^{dW}=T\pi\circ X_H\circ dW,
\end{equation}
where  $T{\pi}:TT^{*}Q\rightarrow T^{*}Q$ is the tangent map of the canonical projection
$\pi:T^{*}Q\rightarrow Q$. The construction of $X_{H}^{dW}$ can be seen on a commutative diagram

\[
\xymatrix{ T^{*}Q
\ar[dd]^{\pi} \ar[rrr]^{X_H}&   & &TT^{*}Q\ar[dd]^{T\pi}\\
  &  & &\\
 Q\ar@/^2pc/[uu]^{dW}\ar[rrr]^{X_H^{dW}}&  & & TQ}
\]
\noindent
Notice that the image of $dW$ is a Lagrangian submanifold, since it is exact and consequently, closed. Indeed, one can change $dW$ by a closed one-form $\gamma$. Lagrangian submanifolds are very important objects in Hamiltonian mechanics, since the dynamical equations (Hamiltonian or Lagrangian)
can be described as Lagrangian submanifolds of convenient (symplectic) manifolds.
We enunciate the following theorem.

\begin{theorem}\label{theorintro}
\label{HJT} For a closed one-form $\gamma=dW$ on $Q$ the following conditions are
equivalent:
\begin{enumerate}
\item The vector fields $X_{H}$ and $X_{H}^{\gamma }$ are $\gamma$-related,
that is
\begin{equation}
T\gamma \circ X_H^{\gamma}=X_H\circ\gamma.
\end{equation}
\item The following equation is fulfilled 
\[
d\left( H\circ \gamma \right)=0.
\]
\end{enumerate}
\end{theorem}

\noindent Theorem \ref{theorintro} is a geometric formulation of the (time-independent) HJ theory from classical mechanics. The first item in the theorem says that if $\left( q^i(t) \right) $ is an integral curve of $X_{H}^{\gamma }$, then $\left(
q^i( t) ,\gamma_j (t)\right) $ is an
integral curve of the Hamiltonian vector field $X_{H}$, hence a solution of the Hamilton equations. Such a solution of the Hamiltonian
equations is called horizontal since it is on the image of a one-form on $Q$. In the local picture, the second condition implies that exterior
derivative of the Hamiltonian function on the image of $\gamma $ is closed,
that is, $H\circ \gamma $ is constant. Under the assumption that $\gamma $ is closed, we can find (at least
locally) a function $W$ on $Q$ satisfying $dW=\gamma $. The pioneer in this purpose was Tulczyjew who characterized the image of local Hamiltonian vector fields on a symplectic manifold $(M,\omega)$ as Lagrangian submanifolds
of a symplectic manifold $(TM,\omega^{T})$, where $\omega^T$ is the tangent lift of $\omega$ to $TM$ \cite{Tulczy}. This result was later generalized to Poisson manifolds \cite{GrabUrb}. In our paper, we generalize this result to time-dependent lcs manifolds and show that the HJ theory is naturally related to canonical transformations in the time-dependent and lcs frame, in similar fashion as in the symplectic case.



The paper is organised as follows. In Section 2, we review the geometric fundamentals of time-dependent Hamiltonian formalism on a symplectic manifold. We recall how the time-dependency is introduced to Hamiltonian mechanics within its main objects: Hamiltonian vector fields, integral curves and Hamilton equations. Next we discuss canonical transformations in the context of Hamiltonian mechanics. Section 3 contains a review of the geometric fundamentals of lcs manifolds and recalls some introductory objects such as musical mappings, the Lichnerowicz-deRham differential,
and the construction of lcs structures on the cotangent bundle. Alongside, we remind the reader of the concept of Lagrangian submanifolds on lcs structures, since these will play a crucial role describing the dynamics on lcs cotangent bundles. 
Section \ref{sec:LCSHam} is the core of our paper. We present the time-dependent Hamiltonian formalism on lcs manifolds, we extend the theory of canonical transformations to the lcs case and subsequently formulate a Hamilton-Jacobi theory on lcs manifolds for explicitly time-dependent Hamiltonians. Section 5 contains one of the possible applications of our formalism, which is concerned with differential equations that admit finite-dimensional Lie algebras of vector fields, i.e., the well-known Lie systems \cite{LS}. We will provide a physical interpretation for these systems out of the construction of time-dependent lcs two-forms with the properties of the Lie algebra that is compatible with the system of differential equations.

\section{Fundamentals on time-dependent Hamiltonian systems}\label{sec:TDH}

\subsection{Time-dependent systems}


Let $Q$ be a smooth manifold. We will denote by $TQ$ and $T^*Q$ the tangent and cotangent bundle on $Q$, respectively. We say that a map 
$$X:\mb R\times Q\to TQ,\qquad (t,q)\longmapsto X(t,q) $$
is a time-dependent vector field if for each $t\in\mb R$, the map
$X_t:Q\to TQ,  q\longmapsto X(t,q)$
is a vector field on $Q$. Therefore, each time-dependent vector field on a manifold can be understood as a family of time-independent vector fields $\{X_t\}_{t\in\mb R}$ that depends in a smooth way on a parameter $t\in\mb R$. 
With each time-dependent vector field we can associate in a unique way a vector field on $\mathbb R\times Q$ given by
$$ \tilde X: \mb R\times Q\to T(\mb R\times Q)\simeq T\mb R\times TQ, \quad (t,q)\longmapsto ((t,1),(q,X(q,t))   $$
so that
$$\tilde X=\ppt +X(t,q).$$
We refer to $\tilde X$ as the suspension or autonomization of the time-dependent vector field $X$. A curve $b:I\to Q$ on a manifold $Q$ is an integral curve of $X$ if 
$$\dot b(t)=X(t,b(t)),\qquad  \forall t\in I, $$
where we used a notation $\dot b(t)=\frac{d}{ds}_{|{s=0}}b(t+s).$

\noindent
Let $(P,\omega)$ be a symplectic manifold, i.e., $P$ is a smooth manifold and $\omega$ a closed and nondegenerate two-form on $P$. We consider the product $\mathbb R\times P$ and the projection $\pi_2:\mathbb R\times P\to P$ onto the second factor. We will denote by $\tilde\omega$ the pull-back of $\omega$ with respect to $\pi_2$, namely $\tilde\omega=\pi_2^*\omega$. A pair $(\mathbb R\times P,\tilde\omega)$ is then a contact manifold with a contact form $\tilde\omega$. The characteristic line bundle of $\tilde\omega$ is generated by the vector field $X=\frac{\partial}{\partial t}$. A time-dependent Hamiltonian is a smooth function $H:\mb R\times P\to\mb R$ such that for each $t\in\mb R$, $H$ defines the function 
$H_t:P\to\mb R$, $H_t(m)=H(t,m).$
From this, we see that a time-dependent Hamiltonian can be viewed as a family of time-independent Hamiltonians $\{H_t\}$ for each $t\in\mathbb R$. Since each element of this family $H_t$ is a smooth function on a symplectic manifold, we can associate with it a Hamiltonian vector field $X_{H_t}$ with respect to $\omega$, namely
\begin{equation}\label{TimeHamEqn}
\omega(\cdot,X_{H_t}) =dH_t.
\end{equation}
By collecting Hamiltonian vector fields $X_{H_t}$ in (\ref{TimeHamEqn}), for each $t\in\mathbb R$ we obtain a family of vector fields $\{X_{H_t}\}$ on $P$. In light of the discussion above, this family defines a time-dependent vector field given by
$$X_H:\mb R\times P\to TP, (t,p)\longmapsto X_{H_t}(p).$$
We will call $X_H$ a time-dependent Hamiltonian vector field of the (time-dependent) Hamiltonian $H$ with respect to the form $\omega$. As usual, we are interested in the integral curves of $X_H$ that represent phase trajectories of the system. It is a matter of straightforward calculations to show that the integral curve $b$ of $X_{H_t}$ is given by the time-dependent Hamilton equations
\begin{align*}
\frac{d}{dt} q^i(b(t))&= \frac{\partial H}{\partial p_i}(t,b(t)),\\
\frac{d}{dt} p_i(b(t))&= -\frac{\partial H}{\partial q^i}(t,b(t)).
\end{align*}

\noindent
Let us recall that in the time-independent case we have $\mathcal L_{X_H}H=0$, where $\mathcal L$ is the Lie derivative. In the time-dependent case we have instead $ \mathcal L_{X_H}H=\frac{\partial H}{\partial t}. $  If $\frac{\partial H}{\partial t}=0$, then all of the above structures reduce to the standard Hamiltonian formalism on a symplectic manifold.

To finish this section let us notice that the existence of the function $H:\mb R\times P\to\mb R$ allows us to define a new two-form
$$ \omega_H=\tilde\omega+dH\wedge dt,  $$
which will be important in our work later on. One can show that $\tilde X_H$, associated with $H$, is the unique vector field satisfying
$$  i_{\tilde X_H}\omega_H=0, \qquad i_{\tilde X_H}dt=1.  $$
Moreover, if $F$ is the flow of $X_H$ then $F^*\omega=\tilde\omega-dH\wedge dt$.
\subsection{Canonical transformations}\label{ssec:cantrans}

In Hamiltonian mechanics, a canonical transformation is a transformation of coordinates that preserves the form of Hamilton equations. The notion of a canonical transformation is an important concept in symplectic geometry itself but it also plays a crucial role in the Hamilton–Jacobi theory (as a useful method for calculating conserved quantities) and in Liouville's theorem in classical statistical mechanics. Let us stress that although a canonical transformation preserves the form of Hamilton equations, it does not need to preserve the form of the Hamiltonian itself. We will briefly recall now the geometric basics of the theory of canonical transformations. The structures presented here will be generalised to the lcs case in Section 4.

Let $(P_1,\omega_1)$ and $(P_2,\omega_2)$ be symplectic manifolds and $(\mathbb R\times P_1,\tilde\omega_1)$, $(\mathbb R\times P_2,\tilde\omega_2)$ the corresponding contact manifolds. A smooth map 
$F:\mathbb R\times P_1\to \mathbb R\times P_2$ is called a canonical transformation if each of the following holds

\begin{description}
 \setlength\itemsep{0.01em}
  \item[{\bf i)}]$F$ is a diffeomorphism 
  \item[{\bf ii)}]  $F$ preserves time, i.e. $F^*t=t$ or, equivalently, the following diagram is commutative
$$\xymatrix{
  \mathbb R\times P_1 \ar[rr]^{F}   \ar[dr]^{pr_{\mathbb R}} & &   \mathbb R\times P_2  \ar[dl]_{pr_{\mathbb R}}  \\
& \mathbb R&
}
$$
  \item[{\bf iii)}] There exists a function $K_F\in C^\infty(\mathbb R\times P_1)$  such that
$$F^*\tilde\omega_2=\omega_{K_F} \qquad where \qquad \omega_{K_F}=\tilde\omega_1+dK_F\wedge dt.$$
\end{description}

\noindent
From now on, we will say that a diffeomorphism $F:\mathbb R\times P_1\to \mathbb R\times P_2$ has the so-called $S$-property if for each $t$, the map $F_t:P_1\to P_2, F_t(p_1):=F(t,p_1)$ is a symplectomorphism. One can show that the map $F$ has the $S$-property if and only if there exists a one-form $\alpha$ on $\mathbb R\times P_1$ such that 
$F^*\tilde\omega_2=\tilde\omega_1+\alpha\wedge dt.$ The most important properties of canonical transformations are expressed in the following theorem.

\begin{theorem}\label{Theorem1}
Assume that conditions $i)$ and $ii)$ hold. Then, $iii)$ is equivalent to each of the following three statements
\begin{description}
 \setlength\itemsep{0.01em}
  \item[{\bf a)}] For each function $H\in C^\infty(\mb R\times P_2)$ there exists a function $K\in C^\infty(\mb R\times P_1)$, such that  
$$  F^*\omega_H=\omega_K $$
  \item[{\bf b)}]  For all $H\in C^\infty(\mb R\times P_2)$  there exists a $K\in C^\infty(\mb R\times P_1)$ such that 
$$F_*\tilde X_K=\tilde X_H$$ 
where $\tilde X_K$ and $\tilde X_H$ are the suspensions of a Hamiltonian vector field $X_K$ and $X_H$ respectively. 
  \item[{\bf c)}] If $\omega_1=d\Theta_1$ and $\omega_2=d\Theta_2$, then there exists a function $K_F$ such that
$$d(F^*\tilde\Theta_2-\Theta_{K_F})=0$$ 
where
$$\tilde\Theta_1= \pi_2^*\Theta_1+dt, \qquad \tilde\Theta_2= \pi_2^*\Theta_2+dt \qquad and \qquad \Theta_{K_F}=\tilde\Theta_1-K_Fdt.  $$
\end{description}
\end{theorem}
Proof of the above theorem may be found for instance in \cite{AbMa78}. We will omit it here since in Section 4 we will prove Theorem \ref{theorcan}, which is a generalisation of the Theorem \ref{Theorem1}. Let us notice, that condition b) implies that $F$ preserves the form of Hamilton equations.

\subsection{Generating functions of canonical transformations} 

It is an interesting fact that each canonical transformation $F:\mb R\times P_1\to\mb R\times P_2$ can be obtained from a real-valued function on $\mathbb R\times P_1$, the so-called generating function of $F$. We will briefly present here basic properties of generating functions, which play a crucial role in a Hamilton-Jacobi theory, which will be the main subject of Section \ref{ssec:lcsHJ}. Let $F:\mb R\times P_1\to\mb R\times P_2$ be a canonical transformation and $\omega_1, \omega_2$ symplectic forms on $P_1$ and $P_2$ respectively. Let us assume that both $P_1$ and $P_2$ are exact i.e. there exist one-forms $\theta_1$ and $\theta_2$ such that $\omega_1=d\theta_1$ and $\omega_2=d\theta_2$. Then, from condition $c)$ in Theorem \ref{Theorem1} we have that $d(F^*\tilde\theta_2-\theta_{K_F})=0,$ which means that locally there exists a function $W:\mb R\times P_1\to\mb R$ such that $F^*\tilde\theta_2-\theta_{K_F}=dW.$ We will call $W$ a generating function for a canonical transformation $F$. Let us stress that the local existence of $W$ is equivalent to $F$ being canonical. Let us denote by $\dot F$ the coefficient of $dt$ in $F^*\pi_2^*\theta_2$. The following theorem provides a relation between generating functions and Theorem \ref{Theorem1}.
\begin{theorem}
If $F:\mb R\times P_1\to\mb R\times P_2$ is a canonical transformation with a local generating function $W:\mb R\times P_1\to\mb R$ then
$$ K_F=\frac{\partial W}{\partial t}-\dot F$$
and for a Hamiltonian $H$ on $\mathbb R\times P_2$
$$ F_*\tilde X_K=\tilde X_H \quad where \quad K=H\circ F+ \Big(\frac{\partial W}{\partial t}-\dot F \Big)  $$ 
where $K$ and $K_F$ are the ones displayed in Theorem \ref{Theorem1}.
\end{theorem}


\section{Geometry of locally conformal symplectic manifolds}

\subsection{Basics on locally conformal symplectic manifolds} \label{blcs}

The pair $(M,\Omega)$ where $\Omega$ is a non-degenerate two-form is called an almost symplectic manifold, and $\Omega$ is an almost symplectic two-form. If $\Omega$ is additionally closed, then the manifold turns out to be a symplectic manifold. There is an intermediate step between symplectic manifolds and almost symplectic manifolds: these are the so called locally conformal symplectic manifolds \cite{vaisman}. An almost symplectic manifold $(M,\Omega)$ is a lcs manifold if the two-form is closed locally up to a conformal parameter, i.e., if there exists an open neighborhood, say $U_\alpha$, around each point $x$ in $M$, and a function $\sigma_{\alpha}$ such that the exterior derivative $d(e^{-\sigma_{\alpha}}\Omega\vert_{\alpha})$ vanishes identically on $U_\alpha$. Here, $\Omega\vert_{\alpha}$ denotes the restriction of the almost symplectic structure $\Omega$ to the open set $U_\alpha$. The positive character of the exponential function implies that the local two-form $e^{-\sigma_{\alpha}}\Omega\vert_{\alpha}$ is non-degenerate as well. Being closed and non-degenerate,
\begin{equation}\label{ua}
    \Omega_{\alpha}=e^{-\sigma_{\alpha}}\Omega\vert_{\alpha}
\end{equation}
is a symplectic two-form. That is, the pair $(U_\alpha, \Omega_{\alpha})$ is a symplectic manifold. 

The question now is how to glue the behavior in all local open charts to arrive at a global definition for lcs manifolds. Notice that, $\Omega\vert_{\alpha}$ is a local realization of the global two-form $\Omega$, whereas, up to now, $\Omega_{\alpha}$ is defined only on $U_\alpha$. In another local chart, say $U_{\beta}$, a local symplectic two-form is defined to be $\Omega_{\beta}:=e^{-\sigma_{\beta}}\Omega\vert_{\beta}$. This gives that, for overlapping charts, the local symplectic two-forms are related by $\Omega_{\beta}=e^{-(\sigma_{\beta}-\sigma_{\alpha})}\Omega_{\alpha}$. Accordingly, this conformal relation determines scalars
\begin{equation} \label{transition}
\lambda_{\beta \alpha}=e^{\sigma_{\alpha}}/e^{\sigma_{\beta}}=e^{-(\sigma_{\beta}-\sigma_{\alpha})}
\end{equation}
 satisfying the cocycle condition 
\begin{equation}\label{cocycle}
    \lambda_{\beta \alpha}\lambda_{\alpha \gamma}=\lambda_{\beta \gamma}.
\end{equation}
This way one can glue the local symplectic two-forms $\Omega_\alpha$ to a line bundle $L \mapsto M$ valued two-form $\tilde{\Omega}$ on $M$. To sum up, we say that there are two global two-forms $\Omega$ (real valued) and $\tilde{\Omega}$ (line bundle valued) on $M$ with local realizations $\Omega\vert_\alpha$ and $\Omega_\alpha$, respectively. These local two-forms are related as in (\ref{ua}). Now, recalling \eqref{ua} once more, it is easy to see that $d\Omega\vert_{\alpha}=d\sigma_{\alpha}\wedge\Omega\vert_{\alpha}$, but equally,
$d\Omega\vert_{\beta}=d\sigma_{\beta}\wedge \Omega\vert_{\beta}$ on an overlapping region $U_\alpha \cap U_\beta$. This implies that $d(\sigma_{\beta}-\sigma_{\alpha})\wedge \Omega|_{U_{\alpha}\cap U_{\beta}}=0,$ and since $\Omega$ is nondegenerate, necessarily, $d\sigma_{\alpha}=d\sigma_{\beta}$. So that, $\theta=d\sigma_{\alpha}$ is a well defined one-form on $M$ that satisfies $d\Omega=\theta\wedge \Omega$. Such a one-form $\theta$ is called the Lee one-form \cite{Hwa}. Since $\theta$ is locally exact, then it is closed.  A lcs manifold $(M,\Omega,\theta)$ is 
a globally conformal symplectic (gcs) manifold if the Lee form $\theta$ is an exact one-form. Since it is fulfilled that in two-dimensional manifolds every closed form is exact, two-dimensional lcs manifolds are gcs manifolds. 
Notice that the Lee form $\theta$ is completely determined by $\Omega$ for manifolds with dimension $4$ or higher. We can likewise denote a lcs manifold by a triple $(M,\Omega,\theta)$. Equivalently, this realization of locally conformal symplectic manifolds reads that a lcs manifold is a symplectic manifold if and only if the Lee form $\theta$ vanishes identically. 
 Conversely, if $(M, \Omega, \theta)$ is a triple such that $\Omega$ is an almost symplectic form, and $\theta$ is a closed one-form such that $d\Omega = \theta \wedge \Omega$, then one can find an open cover $\{U_\alpha\}$ of $M$ such that, on each chart $U_\alpha$, $\theta = d \sigma_\alpha$ for some functions $\sigma_\alpha$. It is clear now that $e^{-\sigma_\alpha} \Omega\vert_\alpha$ is symplectic on $U_\alpha$.\bigskip

\noindent 

\noindent \textbf{Musical mappings.} Consider an almost symplectic manifold $(M,\omega)$. The non-degeneracy of the two-form $\Omega$ leads us to define a musical isomorphism 
\begin{equation} \label{mus-iso}
\Omega^\flat:\mathfrak{X}(M)\longrightarrow \Lambda^1(M): X \mapsto \iota_X\omega,
\end{equation}
where $\iota_X$ is the interior derivative. Here, $\mathfrak{X}(M)$ is the space of vector fields on $M$ whereas $\Lambda^1(M)$ is the space of one-form sections on $M$. 
We denote the inverse of the isomorphism (\ref{mus-iso}) by $\Omega^\sharp$. When pointwise evaluated, the musical mappings $\Omega^\flat$ and $\Omega^\sharp$ induce isomorphisms from $TM$ to $T^*M$, and from $T^*M$ to $TM$, respectively. We shall use the same notation for the induced isomorphisms. 

Let us now concentrate on the particular case of lcs manifolds.  Assume a lcs manifold $(M, \Omega)$ with a Lee form $\theta$. 
Referring to the $\Omega^\sharp$, we define the so called Lee vector field 
\begin{equation} \label{Lee-v-f}
Z_\theta:=\Omega^\sharp(\theta), \qquad \iota_{Z_\theta}\Omega=\theta
\end{equation}
where $\theta$ is the Lee-form. By applying $\iota_{Z_\theta}$ to both sides of the second equation, one obtains that $\iota_{Z_\theta}\theta=0$. Further, by a direct calculation, we see that $\mathcal{L}_{Z_\theta}\theta=0$ and that $\mathcal{L}_{Z_\theta}\Omega=0$. Here, $\mathcal{L}_{Z_\theta}$ is the Lie derivative.
\bigskip
 
\noindent \textbf{The Lichnerowicz-deRham differential.}
Consider now an arbitrary manifold $M$ equipped with a closed one-form $\theta$.  The Lichnerowicz-deRham differential (LdR) on the space of differential forms $\Lambda(M)$ is defined as
\begin{equation} \label{LdR-Diff}
d_\theta: \Lambda^k(M) \rightarrow\Lambda^{k+1}(M) : \beta \mapsto d\beta-\theta\wedge\beta,
\end{equation}
where $d$ denotes the exterior (deRham) derivative. Notice that $d_\theta$ is a differential operator of order $1$. That is, if $\beta$ is a $k$-form then $d_\theta\beta$ is $k+1$-form. The closure of the one-form $\theta$ reads that $d_{\theta}^2=0$. This allows the definition of  cohomology as the $d_\theta$ cohomology in $\Lambda(M)$
\cite{HaRy99}. We represent this with the pair $(\Lambda(M),d_{\theta})$. 
A direct computation shows that an almost symplectic manifold $(M, \Omega)$ equipped with a closed one-form $\theta$ is a lcs manifold if and only if $d_\theta \Omega = 0$. 
\bigskip

\noindent \textbf{{Lagrangian Submanifolds of lcs manifolds.} }
Consider an almost symplectic manifold $(M, \Omega)$. Let $L$ be a submanifold of $M$.
The complement $TL^{\bot}$ is defined with respect to $\Omega$. For a point $x\in L$,
\begin{equation}\label{lagsub}
    T_{x}L^{\bot}=\{u\in T_xM \enskip|\enskip \Omega(u,w)=0, \forall w\in T_xL\}.
\end{equation}
\noindent 
We say that $L$ is isotropic if $TL\subset TL^{\bot}$, it is coisotropic if $TL^{\bot}\subset TL$ and it is Lagrangian if $TL^{\bot}=TL$. Accordingly a submanifold is Lagrangian if it is both isotropic and coisotropic. Observe that the definition is exactly the same as in the symplectic case, since they are obtained at the linear level.

\subsection{Locally conformal symplectic structures on cotangent bundles} \label{Sec-lcs-cot}
We shall depict the lcs framework on cotangent bundles. Start with the canonical symplectic manifold $(T^*Q,\omega_Q)$. Here, the canonical symplectic two-form $\omega_Q = - d \Theta_Q$ is minus the exterior derivative of the canonical Liouville one-form $\Theta_Q$  on $T^*Q$. Let $\vartheta$  be a closed one-form on the base manifold $Q$ and pull it back to $T^*Q$ by means of the cotangent bundle projection $\pi_Q$. This gives us a closed semi-basic one-form $\theta=\pi_Q^*(\vartheta)$. By means of the Lichnerowicz-deRham differential, we define a two-form 
\begin{equation} \label{omega_theta}
\Omega_\theta=-d_\theta(\Theta_Q)= -d\Theta_Q+\theta\wedge \Theta_Q=\omega_Q+\theta\wedge \Theta_Q
\end{equation}
on the cotangent bundle $T^*Q$. Since $d\Omega_\theta=\theta\wedge \Omega_\theta$ holds, the triple 
\begin{equation} \label{T*_Q}
T^*_\theta Q=(T^*Q,\Omega_\theta,\theta)
\end{equation} 
determines a locally conformal symplectic manifold with Lee-form $\theta$. In short, we denote this lcs manifold by simply $T^*_\theta Q$. This structure is
conformally equivalent to a symplectic manifold if and only if $\vartheta$ lies in the zeroth class of the first deRham cohomology on $Q$. Notice that $T^*_\theta Q$ is an exact locally conformal symplectic manifold since $\Omega_\theta$ is defined to be minus of the Lichnerowicz-deRham differential $d_\theta$ of the canonical one-form $\Theta_Q$. It is important to note that all lcs manifolds locally look like $T^*_\theta Q$ for some $Q$ and for a closed one-form $\vartheta$. 
\bigskip

\noindent 
Consider the lcs manifold $T_\theta^*Q$ in (\ref{T*_Q}) with Lee form $\theta=\pi_Q^*\vartheta$. Let $\gamma$ be a section of the cotangent bundle or, in other words, a one-form on $Q$. A direct computation shows that the pull-back of the lcs structure is $d_\theta$ exact, that is 
\begin{equation} \label{LagSubT*Q}
\gamma^* \Omega_\theta = - d_\vartheta \gamma
\end{equation} 
where $d_\vartheta$ denotes the LdR differential defined by the one-form $\vartheta$ on $Q$. This implies that the image space of $\gamma$ is a Lagrangian submanifold of $T_\theta^*Q$ if and only if $d_\vartheta \gamma=0$. Since $d_\vartheta^2$ is identically zero, the image space of the one-form $d_\vartheta f$ is a
 Lagrangian submanifold of $T^*_\theta Q$ for some function $f$ defined on $Q$. 
\subsection{Dynamics on locally conformal symplectic manifolds }

Let us now concentrate on Hamiltonian dynamics on lcs manifolds \cite{vaisman13}.  As discussed previously, there are two equivalent definitions of lcs manifolds. One is local, and the other is global. First consider the local definition by recalling the local symplectic manifold $(U_\alpha,\Omega_\alpha)$. For a Hamiltonian function $h_\alpha$ on this chart, we write the Hamilton equations by  
\begin{equation}\label{geohamalpha}
    \iota_{X_{\alpha}}\Omega_{\alpha}=dh_{\alpha}.
\end{equation}
Here, $X_{\alpha}$ is the local Hamiltonian function associated to this framework. In terms of the Darboux coordinates $(q^i_{(\alpha)},p_i^{(\alpha)})$ on $U_\alpha$. The local symplectic two-form is $\Omega_{\alpha}=dq^i_{(\alpha)}\wedge dp_{i}^{(\alpha)}$, and the Hamilton equation (\ref{geohamalpha}) becomes  
\begin{equation} \label{HamEqLoc}
    \frac{dq^i_{\alpha}}{dt}=\frac{\partial h_{\alpha}}{\partial p_i^{\alpha}},\qquad \frac{dp_i^{\alpha}}{dt}=-\frac{\partial h_{\alpha}}{\partial q_{\alpha}^i}.
\end{equation}
We have discussed the gluing problem of the local symplectic manifolds but we have not addressed this problem for the local Hamiltonian functions. We wish to define a global realization of the local Hamiltonian functions in such a way that the structure of the local Hamilton equations (\ref{HamEqLoc}) does not change under transformations of coordinates. This can be rephrased as to establish a global realization of the local Hamiltonian function $h_\alpha$ by preserving the local Hamiltonian vector fields $X_\alpha$ in (\ref{geohamalpha}). A direct observation reads that multiplying both sides of (\ref{geohamalpha}) by the scalars $\lambda_{\beta\alpha}$ defined in (\ref{transition}) leaves the dynamics invariant. So that, the transition 
$h_{\beta}=e^{\sigma_{\alpha}-\sigma_{\beta}}h_{\alpha}$ is needed for the preservation of the structure of the equations. In the light of the cocyle character of the scalars shown in (\ref{cocycle}), we can glue the local Hamiltonian functions $h_\alpha$ to define a section $\tilde{h}$ of the line bundle $L\mapsto M$. On the other hand, in the light of the identity $e^{\sigma_{\alpha}}h_{\alpha}=e^{\sigma_{\beta}}h_{\beta}$ one arrives at a real valued Hamiltonian function 
\begin{equation} \label{glueHamFunc}
h\vert_\alpha=e^{\sigma_\alpha}h_\alpha.
\end{equation}
on $U_\alpha$ that defines a real valued function $h$ on the whole $M$. Recall the discussion in Subsection \ref{blcs} about the local and global character of the two-forms. Similarly, we argue that there exist two global functions $\tilde{h}$ (line bundle valued) and $h$ (real valued) on the manifold $M$. On a chart $U_\alpha$, these functions reduce to $h_\alpha$ and $h\vert_\alpha$, respectively, and they satisfy relation \eqref{glueHamFunc}.

In order to recast the global picture of the Hamilton equation (\ref{geohamalpha}), we first substitute identity (\ref{glueHamFunc}) into \eqref{geohamalpha}. Hence, a direct calculation turns the Hamilton equations into the following form
\begin{equation} \label{LocHam2}
\iota_{X_\alpha} \Omega\vert_\alpha =dh\vert_\alpha-h\vert_\alpha d\sigma_\alpha
\end{equation}
where we have employed the identity (\ref{ua}) on the left hand side of this equation. Notice that all the terms in equation (\ref{LocHam2}) have global realizations. So, we can write
\begin{equation}\label{semiglobal}
   \iota_{X_{h}}\Omega=dh-h\theta,
\end{equation}
where $X_{h}$ is the vector field obtained by gluing all the vector fields $X_\alpha$. That is, we have $X_h\vert_\alpha=X_\alpha$. Notice that \eqref{semiglobal} can also be written as 
\begin{equation}\label{semiglobal3}
\iota_{X_h} \Omega = d_\theta h,
\end{equation}
where $d_\theta$ is the Lichnerowicz-deRham differential given in (\ref{LdR-Diff}). The vector field $X_h$ defined in (\ref{semiglobal3}) is called a Hamiltonian vector field for the Hamiltonian function $h$. In terms of the Lee vector field $Z_\theta$ defined in (\ref{Lee-v-f}), the Hamiltonian vector field is computed to be 
\begin{equation}\label{vflcs}
X_h=\Omega^\sharp(dh)+hZ_\theta,
\end{equation}
where $\Omega^\sharp$ is the musical isomorphism induced by the almost symplectic two-form $\Omega$. From this, one can easily see that, apart from the classical symplectic framework, for the constant function $h = 1$ the corresponding Hamiltonian vector field is not zero but the Lee vector in (\ref{Lee-v-f}), that is $Z_\theta=X_1$. More generally, a vector field $X$ is called a locally conformal Hamiltonian vector field if 
\begin{equation}
d _ { \theta } ( \iota_{X}\Omega)=0.
\end{equation}
It is immediate to see that a Hamiltonian vector field is a locally Hamiltonian vector field since $d^2_\theta =0$.  

\subsection{Morphisms of lcs manifolds}

Let $(M_1,\Omega_1,\theta_1)$ and $(M_2,\Omega_2,\theta_2)$ be lcs manifolds of the same dimension. We say that the diffeomorphism $F:M_1\to M_2$ is a morphism of lcs manifolds if $F^*\Omega_2=\Omega_1.$ Notice that this condition implies that
$$ d\Omega_1=d(F^*\Omega_2)=F^*d\Omega_2=F^*\theta_2\wedge F^*\Omega_2=F^*\theta_2\wedge\Omega_1,  $$
which means that $F^*\theta_2=\theta_1.$ In a local picture we have that if $\theta_1=d\sigma_1$ and $\theta_2=d\sigma_2$ for some $\sigma_1:M_1\to\mathbb R$ and $\sigma_2:M_2\to\mathbb R$ then $F\circ\sigma_2=\sigma_1$. From now on we will call $F$ a morphism of locally conformal symplectic manifolds or in short a lcs-morphism. 



\section{Time-dependent Hamiltonian formalism on a lcs framework}\label{sec:LCSHam}


In this section we present the main result of our paper, which consists of the formulation of time-dependent Hamiltonian mechanics on a lcs manifold and the subsequent discussion of canonical transformations and Hamilton-Jacobi theory in the context of lcs manifolds. 

In \cite{ChLeMa91} one can find a brief discussion of lcs structures in relation to a cosymplectic framework. In our approach we prefer a more direct and physical approach in the spirit of \cite{AbMa78} starting directly from a phase space of the form $\mathbb R\times M$, where $M$ is a lcs manifold. In physical applications $\mathbb R\times M$ represents the phase space of a time-dependent system on a lcs manifold. Subsequently, we present a generalization of the theory of canonical transformations to a lcs structure. We discuss it in both the local and the global picture and show how this theory naturally arises as an extension of the cosymplectic case. We finish our study by setting a Hamilton-Jacobi theory in the context of time-dependent lcs manifolds. For review of the time-independent case we refer the reader to \cite{EsLeZajSard}. 


\subsection{Time-dependent Hamiltonian dynamics in a lcs framework}\label{ssec:LCSTimeHam}

The time-dependent Hamiltonian formalism on a lcs manifold may be constructed from the time-independent one in similar fashion as it is performed in the symplectic case. Since locally a lcs manifold resembles a symplectic manifold, we can use the results from Section \ref{sec:TDH} for each chart $U_\alpha$ in $M$ and find the transition conditions on double overlaps $U_\alpha\cap U_\beta\subset M$.

\noindent
Let $(M,\Omega,\theta)$ be a lcs manifold. Locally we have a family of open charts $U_\alpha$ and functions $\sigma_\alpha:U_\alpha\to\mathbb R$. Let us recall that each $(U_\alpha,\Omega_\alpha)$, where $\Omega_\alpha=e^{-\sigma_\alpha}\Omega |{_\alpha}$, is a symplectic manifold, which means that we can construct a time-dependent Hamiltonian structure on it, applying the constructions from Section \ref{sec:TDH}. Let $H_{t\alpha}:\mathbb R\times U_\alpha\to\mathbb R$ be a time-dependent Hamiltonian function for a given $t\in\mathbb R$. Then, according to (\ref{TimeHamEqn}), the dynamics on a local chart $U_\alpha$ is given by the equation $\iota_{X_{t\alpha}}\Omega_{\alpha}=dH_{t\alpha},$ where $X_{t\alpha}$ is a time-dependent vector field representing the dynamics on $U_\alpha$. The global time-dependent Hamiltonian $H$ may be obtained from gluing local time-dependent Hamiltonians on $U_\alpha\cap U_\beta$ through the condition $e^{\sigma_\alpha}H_{t\alpha}=e^{\sigma_\beta}H_{t\beta}$, for each  $t\in\mathbb R$. Therefore, we obtain a global Hamiltonian function 
$$ H:\mathbb R\times M\to\mathbb R, \qquad H_t|{_\alpha}=e^{\sigma_\alpha}H_{t\alpha},  $$
where $H_t(m):=H(t,m)$. We have the diagram
{\large
$$
\xymatrix{ 
M\times\mathbb R  \ar[drr]_{pr_{\mathbb R}}  \ar[d]^{pr_{M}}  \ar@/^1pc/[drr]^{H}   &&   \\ 
M \ar@/_1pc/[rr]_{H_t}  && \mathbb R
}
$$
}
\noindent The dynamics of the system is given by the equation
\begin{equation}\label{lcshamdyn}
\iota_{X_{H_t}}\Omega=d_\theta H_t,
\end{equation}
where $X_{H_t}$ is a Hamiltonian vector field on $M$ associated with the Hamiltonian $H_t$. 
Equation (\ref{lcshamdyn}) constitutes a generalization of the time-dependent Hamilton equation (\ref{TimeHamEqn}) to a lcs framework. The family of vector fields $\{X_{H_t} \}$ defines a time-dependent vector field 
$$X_H:\mb R\times M\to TM,\quad  (t,m)\longmapsto X_{H_t}(m),$$
which represents the phase dynamics of the system, where the phase space is represented by $\mathbb R\times M$. The solutions of $X_H$ are integral curves $b:\mathbb R\to M$ of $X_H$, that represent the phase trajectory of the system. The equations for the integral curves $b:\mathbb R\to M$ read
\begin{eqnarray}\label{tHameq}
\frac{d}{dt} q^i(b(t))&=& \frac{\partial H}{\partial p_i}(t,b(t)),   \nonumber\\
\frac{d}{dt} p_i(b(t))&=& -\frac{\partial H}{\partial q^i}(t,b(t))+\frac{\partial H}{\partial p_k}(t,b(t))(\theta_kp_i-p_k\theta_i)+\theta^iH.
\end{eqnarray}
\noindent Let us notice that for $\theta=0$, the above equations reduce to standard time-dependent Hamilton equations.

In the following sections we will need the pull-backs of objects defined on $M$ by means of the projection $pr_2:\mathbb R\times M\to M$. For a lcs manifold  $(M_,\Omega,\theta)$ we define the pull-backs $ \tilde\Omega:=pr_2^*\Omega,$ $\tilde\theta:=pr_2^*\theta.$
See that locally we have $\theta|_\alpha=d\sigma_\alpha$, so that
$$ \tilde\theta|_\alpha=pr_2^*d\sigma_\alpha=dpr_2^*\sigma_\alpha=d\sigma_\alpha\circ pr_2=d\tilde\sigma_\alpha,  $$
where we have used the notation $ \tilde\sigma_\alpha:\mathbb R\times U_\alpha\to\mathbb R$, $\tilde\sigma_\alpha=\sigma_\alpha\circ pr_2.$ Furthermore, for an exact $\Omega$, i.e. $\Omega=d_\theta\Theta$, we obtain
$$ \tilde\Omega=pr^*_2d_\theta\Theta=d_{pr^*_2\theta}pr^*_2\Theta=d_{\tilde\theta}\tilde\Theta.  $$

\noindent
At the end of this section we will point out that, similarly to the symplectic case, the existence of the function $H:\mb R\times M\to\mb R$ allows us to define a new two-form
\begin{equation}
 \Omega_{H}=\tilde\Omega+d_{\tilde\theta}H\wedge dt,
\end{equation}
which will be important in our work later on. One can show that $\tilde X_H$, associated with $H$, is the unique vector field satisfying
\begin{equation}\label{lcscontact}
  i_{\tilde X_H}\Omega_{H}=0, \qquad i_{\tilde X_H}dt=1.
\end{equation}
It is a matter of straightforward computation to show that the equation for $X_H$ associated with $\tilde X_H$ coming from (\ref{lcscontact}) takes the form (\ref{tHameq}) in local coordinates.

\subsection{Canonical transformations on lcs manifolds}

We will present now one of the main results of this paper, which is the notion of canonical transformations on lcs manifolds. Before we arrive at the global definition, we will take a look at the local picture. Let $(M_1,\Omega_1,\theta_1)$ and $(M_2,\Omega_2,\theta_2)$ be lcs manifolds. Let us choose open subsets $U_1\subset M_1$ and $U_2\subset M_2$. Then $(U_{1\alpha},\Omega_{1\alpha},\sigma_{1\alpha})$ and $(U_{2\alpha},\Omega_{2\alpha},\sigma_{2\alpha})$ are by definition symplectic manifolds. We can consider a canonical transformation between these two symplectic manifolds i.e. a diffeomorphism $F_\alpha:\mathbb R\times U_{1\alpha}\to\mathbb R\times U_{2\alpha}$ satisfying 
$$ F_\alpha^*\tilde\Omega_{2\alpha}=\tilde\Omega_{1\alpha}+dK_\alpha\wedge dt,  $$ 
where $\tilde\Omega_1|_\alpha$ and $\tilde\Omega_2|_\alpha$ are pull-backs of the forms $\Omega_1|_\alpha$ and $\Omega_2|_\alpha$ with respect to the projections $\mathbb R\times U_{1\alpha}\to U_{1\alpha}$ and $\mathbb R\times U_{2\alpha}\to U_{2\alpha},$ respectively and $K_\alpha$ is a function on $\mathbb R\times U_{1\alpha}$. By definition of the forms $\Omega_1$ and $\Omega_2$ we have that $\tilde\Omega_{1\alpha}=e^{-\tilde\sigma_{1\alpha}}\tilde\Omega_1|_\alpha$ and  $\tilde\Omega_{2\alpha}=e^{-\tilde\sigma_{2\alpha}}\tilde\Omega_2|_\alpha$ so we can write
$$ F_\alpha^*e^{-\tilde\sigma_{2\alpha}}\tilde\Omega_2|_\alpha=e^{-\tilde\sigma_{1\alpha}}\tilde\Omega_1|_\alpha+dK_\alpha\wedge dt.  $$ 
The global definition of a canonical transformation on a lcs manifolds should come from the gluing of local canonical transformations on charts of $M_1$ and $M_2$. Therefore, let us multiply the above formula by $e^{\tilde\sigma_{1\alpha}}$ so that
\begin{equation}\label{equatonF}
e^{\tilde\sigma_{1\alpha}}F_\alpha^*e^{-\tilde\sigma_{2\alpha }}\tilde\Omega_2|_\alpha=\tilde\Omega_1|_\alpha+e^{\tilde\sigma_{1\alpha}}dK_\alpha\wedge dt.
\end{equation}
Now we can make a few observations. First of all, let us assume that 
\begin{equation}\label{conditFsigma}
 \tilde\sigma_{1\alpha}=\tilde\sigma_{2\alpha }\circ F_\alpha. 
\end{equation}
Then we can write 
$$ e^{\tilde\sigma_{1\alpha}}F_\alpha^*e^{-\tilde\sigma_{2\alpha }}\tilde\Omega_2|_\alpha=  e^{\tilde\sigma_{1\alpha}}e^{-\tilde\sigma_{2\alpha }\circ F_\alpha}F_\alpha^*\tilde\Omega_2|_\alpha=F_\alpha^*\tilde\Omega_2|_\alpha, $$
where the last equality comes from (\ref{conditFsigma}). On the other hand, it is easy to check that  $e^{\tilde\sigma_{1\alpha}}dK_\alpha=d_{d\tilde\sigma_{1\alpha}}(e^{\tilde\sigma_{1\alpha}}K_\alpha)$, where $d_{d\tilde\sigma_{1\alpha}}$ is the Lichnerowicz-deRham differential with respect to a form $d\tilde\sigma_{1\alpha}$. Locally we have $\tilde\theta_1|_{\alpha}=d\tilde\sigma_{1\alpha}$ so finally we can rewrite (\ref{equatonF}) as
$$ F_\alpha^*\tilde\Omega_2|_\alpha=\tilde\Omega_1|_\alpha+d_{\tilde\theta_1|_{\alpha}}(e^{\tilde\sigma_{1\alpha}}K_\alpha)\wedge dt.  $$ 
From the above equation it is easy to see the global version of the equation, which reads
$$ F^*\tilde\Omega_2=\tilde\Omega_1+d_{\tilde\theta_1}K\wedge dt,  $$ 
where $F|_\alpha=F_\alpha$ and $K$ is a function on $\mathbb R\times M_1$, such that locally $K|_\alpha= e^{\tilde\sigma_{1\alpha}}K_\alpha$. Notice that by imposing condition (\ref{conditFsigma}) chart by chart we obtain the global condition $F^*\tilde\theta_2=\tilde\theta_1$. Furthermore, it is obvious that $F$ preserves time i.e. $F^*t=t$.



\noindent
Considering the above conclusions, we arrive at the following global definition of a canonical transformation on a lcs manifold.

\begin{definition}
Let $(M_1,\Omega_1,\theta_1)$ and $(M_2,\Omega_2,\theta_2)$ be lcs manifolds and $ \tilde\Omega_1=pr^*_2\Omega_1,  \tilde\Omega_2=pr^*_2\Omega_2,$ be the corresponding two-forms on $\mathbb R\times M_1$ and $\mathbb R\times M_2$, respectively. A smooth map 
$$F:\mathbb R\times M_1\to \mathbb R\times M_2$$
is called a canonical transformation if the following statements hold
\begin{description}
 \setlength\itemsep{0.01em}
  \item[{\bf i)}]$F$ is a diffeomorphism 
  \item[{\bf ii)}]  $F$ preserves time, i.e. $F^*t=t$ or, equivalently, the following diagram is commutative
$$\xymatrix{
  \mathbb R\times M_1 \ar[rr]^{F}   \ar[dr]^{pr_{\mathbb R}} & &   \mathbb R\times M_2  \ar[dl]_{pr_{\mathbb R}}  \\
& \mathbb R&
}
$$
 \item[{\bf iii)}] $F$ satisfies $F^*\tilde\theta_2=\tilde\theta_1$  
\item[{\bf iv)}] There exists a function $K_F\in\mathcal F(\mathbb R\times M_1)$  such that
$$F^*\tilde\Omega_2=\Omega_{K_F} \qquad where \qquad \Omega_{K_F}=\tilde\Omega_1+d_{\tilde\theta_1}K_F\wedge dt.$$
\end{description}

\end{definition}

\noindent
Notice that in comparison with the definition of canonical transformations given in Subsection \ref{ssec:cantrans}, we have an extra condition stating that $F^*\tilde\theta_2=\tilde\theta_1$. The map $F$ reduced to a subset $U_{1\alpha}\subset M_1$ is a canonical transformation between the symplectic manifolds $(U_{1\alpha},\Omega_{1\alpha})$ and $(U_{2\alpha},\Omega_{2\alpha})$, where $U_{2\alpha}:=F(U_{1\alpha})$. Condition iii) ensures that these local canonical transformations glue-up to a global map that establishes a canonical transformation from $M_1$ to $M_2$. 

The following theorem defines the concept of canonical transformations for Hamiltonian mechanics on lcs manifold.

\begin{theorem}\label{theorcan}
Let $i,ii,iii$ hold. Then, the following conditions are equivalent. 
\begin{description}
 \setlength\itemsep{0.01em}
  \item[0.] The condition iv) is satisfied.
  \item[1.] For each function $H\in C^\infty(\mb R\times M_2)$ there exists a function $K\in C^\infty(\mb R\times M_1)$, such that  
$$  F^*\Omega_H=\Omega_K, \quad where \quad  \Omega_H=\tilde\Omega+d_{\tilde\theta_2}H\wedge dt $$  
\item[2]  For each $H\in C^\infty(\mb R\times M_2)$  there exists a $K\in C^\infty(\mb R\times M_1)$ such that 
$$F_*\tilde X_K=\tilde X_H,$$ 
\noindent where $\tilde X_K$ and $\tilde X_H$ is a suspension of a Hamiltonian vector field $X_K$ and $X_H$, respectively. 
  \item[3]  If $\Omega_1=d_{\theta_1}\Theta_1$ and $\Omega_2=d_{\theta_2}\Theta_2$, then there exists a function $K_F$ such that
$$d_{\tilde\theta_1}(F^*\tilde\Theta_2-\Theta_{K_F})=0$$ 
where
$$\tilde\Theta_1= \pi_2^*\Theta_1+dt, \qquad \tilde\Theta_2= \pi_2^*\Theta_2+dt \qquad and \qquad \Theta_{K_F}=\tilde\Theta_1-K_Fdt.  $$  
\end{description} 
\end{theorem}

Proof: \\ \\
$0 \Longleftrightarrow 1. $

Let us take $K=H\circ F+K_F$, then:
$$  F^*\Omega_H=F^*(\tilde\Omega_2+d_{\tilde\theta_2}H\wedge dt)= F^*\tilde\Omega_2+  F^*(d_{\tilde\theta_2}H)\wedge F^*dt=  F^*\tilde\Omega_2+ d_{\tilde\theta_1}(H\circ F)\wedge dt= $$
$$= \tilde\Omega_1+d_{\tilde\theta_1}K_F\wedge dt+ d_{\tilde\theta_1}(H\circ F)\wedge dt= \tilde\Omega_1+d_{\tilde\theta_1}(K_F+H\circ F)\wedge dt=\Omega_K  $$
The converse implication is trivial since by taking $H=0$ and $K=K_F$ we obtain exactly iv).  \\ \\
$1 \Longleftrightarrow 2. $ 

Again we have $K=H\circ F+K_F$. We show first that 1 implies 2. We know that $\tilde X_H$ is the unique vector field satisfying  
$$  i_{\tilde X_H}\Omega_H=0, \qquad i_{\tilde X_H}dt=1.  $$
Therefore, we have to prove that 
$$  i_{(F^{-1})_*\tilde X_H}\Omega_K=0, \qquad i_{(F^{-1})_*\tilde X_H}dt=1.  $$
Indeed, we have
$$  i_{(F^{-1})_*\tilde X_H}\Omega_K=i_{(F^{-1})_*\tilde X_H}F^*\Omega_H=  F^*i_{X_H}\Omega_H=0.$$
and 
$$  i_{(F^{-1})_*\tilde X_H}dt=i_{(F^{-1})_*\tilde X_H}F^*dt= F^*i_{ X_H}dt= F^*1=1,   $$
which proves $(F^{-1})_*\tilde X_H=\tilde X_K$. By reversing arguments it is straightforward 
to prove that 2 implies 1. \\ \\ 
$ 0 \Longleftrightarrow 3. $

From condition iv) we have $F^*\tilde\Omega_2-\Omega_{K_F}=0$. Moreover, we have
$$F^*\tilde\Omega_2-\Omega_{K_F}=F^*\tilde\Omega_2-\tilde\Omega_1-d_{\tilde\theta_1}K_F\wedge dt=F^*d_{\tilde\theta_2}\tilde\Theta_2- d_{\tilde\theta_1}\tilde\Theta_1-d_{\tilde\theta_1}K_F\wedge dt =$$
$$= d_{\tilde\theta_1}(F^*\tilde\Theta_2- \tilde\Theta_1- K_F dt),  $$
so that
$$  F^*\tilde\Omega_2-\Omega_{K_F}=0   \qquad \Longleftrightarrow \qquad  d_{\tilde\theta_1}(F^*\tilde\Theta_2- \tilde\Theta_1- K_F dt)=0.  $$

$\qquad\qquad\qquad\qquad\qquad\qquad\qquad\qquad\qquad\qquad\qquad\qquad\qquad\qquad\qquad\qquad\qquad\qquad\qquad \blacksquare$

\noindent Notice that Theorem \ref{Theorem1} is a special case of Theorem \ref{theorcan}, when $\theta_1=\theta_2=0$. Condition 2 in \ref{theorcan} implies that $F$ preserves the form of generalized Hamilton equations on lcs manifold given by \ref{tHameq}. On the other hand, condition 3 leads us to the notion of generating functions for canonical transformations. Indeed, from
$$d_{\theta_1}(F^*\tilde\Theta_2- \tilde\Theta_1- K_F dt)=0,  $$
we immediately obtain
$$  F^*\tilde\Theta_2- \tilde\Theta_1- K_F dt = d_{\theta_1}W,  $$
where $W:\mathbb R\times M_1\to\mathbb R$. We will call the function $W$ a generating function of a canonical transformation $F$.




\subsection{Hamilton-Jacobi theory on lcs manifolds.}\label{ssec:lcsHJ}

We will present now an extension of the HJ theory to the time-dependent lcs framework. The standard HJ theory is usually related to a symplectic or cosymplectic structure (see e.g. \cite{LeSa}). A time-independent HJ theory for lcs manifolds can be found for instance in \cite{EsLeZajSard}. Here we present its time-dependent version, which can be seen as a generalization containing both approaches from  \cite{EsLeZajSard} and \cite{LeSa}. Since we are interested in mechanics we will restrict ourselves to the case $M=T^*_\theta Q$. The generalisation of the results below to the general case where $M$ is $\theta$-exact and fibered over $Q$ is straightforward.

Consider the fibration $\pi:T^*_\theta Q\times\mathbb R\to Q\times\mathbb R$ and a section $\gamma$ of $\pi:T^*_\theta Q\times\mathbb R\to Q\times\mathbb R$, which preserves time. In local coordinates $\gamma$ can be written as
$$  \gamma: Q\times\mathbb R\to T^*_\theta Q\times\mathbb R, \qquad \gamma(q^i,t)=(q^i,\gamma^i(q^i,t), t). $$
We can associate with $\gamma$ a section $\gamma_t:Q\to T^*_\theta Q\times\mathbb R$, such that $\gamma_t(q^i):=\gamma(q^i,t)$. We have the diagram
$$
\xymatrix{ 
T^*_\theta Q\times\mathbb R  \ar[ddrrr]_{pr_{\mathbb R}}  \ar[dd]^{\pi}    &&&   \\ 
&&  \\
Q\times\mathbb R \ar@/^1pc/[uu]^{\gamma}   \ar[rrr]^{pr_{Q}} &&&   Q  \ar@/_1pc/[uulll]_{\gamma_t}
}
$$
Similarly, we can define a section $\gamma_q(t):=pr_{T^*_\theta Q}\circ\gamma(q,t)$ represented by the diagram
$$
\xymatrix{ 
T^*_\theta Q\times\mathbb R  \ar[rrr]^{pr_{T^*_\theta Q}}  \ar[dd]^{\pi}    &&&  T^*_\theta Q  \\ 
&&  \\
Q\times\mathbb R \ar@/^1pc/[uu]^{\gamma}    \ar@/_1pc/[uurrr]_{\gamma_q} &&&    
}
$$
In the following we will assume that $\gamma$ is $\theta$-closed, i.e. $d_\theta\gamma = 0$. Consider a Hamiltonian vector field $\wdt X_H$ defined through equation (\ref{lcshamdyn}). We can use a section $\gamma$ to define a projected vector field on $Q\times\mathbb R$ as
$$  \wdt X^\gamma_H= T\pi\circ \wdt X_H\circ\gamma.  $$
The following diagram summarizes the above construction
\[
\xymatrix{ T^*_\theta Q\times\mathbb R
\ar[dd]^{\pi} \ar[rrr]^{\tilde X_H}&   & &T(T^*_\theta Q\times\mathbb R)\ar[dd]^{T\pi}\\
  &  & &\\
 Q\times\mathbb R\ar@/^2pc/[uu]^{\gamma}\ar[rrr]^{X_H^{\gamma}}&  & & T(Q\times\mathbb R)}
\]
The last necessary tool for the construction of a HJ theory is the vertical lift of a one-form \cite{IY}. Let $\alpha$ be a one form on $T^*_\theta Q$, which in local coordinates reads $\alpha=\alpha_idq^i$. We say that a vector field $\alpha^V$ is a vertical lift of $\alpha$ if 
$$ i_{\alpha^V}\Omega_\theta=\alpha. $$ 
In local coordinates we have $ \alpha^V=-\alpha_i\frac{\partial}{\partial p_i}$. Since $\Omega_\theta$ is nondegenerate, the map $\alpha\to\alpha^V$ is an isomorphism. 
The following theorem constitutes a time-dependent Hamilton-Jacobi theory for lcs manifolds. 

\begin{theorem}
Let $M=T_\theta^*Q$ be a lcs manifold and let $\wdt X_H$ be a time-dependent Hamiltonian vector field on $\mathbb R\times T_\theta^*Q$ associated with the Hamiltonian $H:\mathbb R\times T_\theta^*Q\to\mathbb R$. Consider a section $\gamma:Q\times\mathbb R\to T_\theta^*Q\times\mathbb R$, such that $d_\theta\gamma = 0$. Then, the following conditions are equivalent:

i) The two vector fields $\tilde X_H$ and $\tilde X^\gamma_H$ are $\gamma$-related i.e. 
$$ T\gamma(\wdt X^\gamma_H)=\wdt X_H\circ\gamma , \qquad where \qquad \wdt X^\gamma_H=T\pi\circ\wdt X_H\circ\gamma.    $$ 

ii) The following equation is fulfilled
\begin{equation}\label{tlcsHJ}
[d(H\circ\gamma_t)+\dot\gamma_q]^V=0.
 \end{equation}
\end{theorem} 
\newpage
\noindent Proof: \\ \\
\noindent We know that $X_H$ is a unique vector field satisfying $i_{X_H}\Omega_{\theta H}=0, i_{X_H}dt=1,$ where $\Omega_{\theta H}=\Omega_{\theta}+d_\theta H\wedge dt$. It is a matter of calculation to check that for
$$ \Omega_{\theta}=dq^i\wedge dp_i+\theta_ip_jdq^i\wedge dq^j ,\qquad \tilde X_H= \frac{\partial}{\partial t}+a^i\frac{\partial}{\partial q^i}+b^j\frac{\partial}{\partial p_j}  $$
one obtains
$$  a^i=\frac{\partial H}{\partial p_i} , \qquad  b^i= -\frac{\partial H}{\partial q^i}+\frac{\partial H}{\partial p_j}(\theta_jp_i-\theta_ip_j)+\theta_iH,   $$
so that
$$ \wdt X_H= \frac{\partial}{\partial t}+\frac{\partial H}{\partial p_i}\frac{\partial}{\partial q^i}+\Big( -\frac{\partial H}{\partial q^i}+\frac{\partial H}{\partial p_j}(\theta_jp_i-\theta_ip_j)+\theta_iH \Big)\frac{\partial}{\partial p_i}.  $$
Now, we take a curve $\gamma:Q\times\mathbb R\to T^*_{\theta}Q\times\mathbb R$, $\gamma(t,q^i)=(t,q^i,\gamma^k(t,q^i))$,
which satisfies $d_\theta\gamma=0$. In coordinates we have
\begin{equation}\label{dgammazero}
d_\theta\gamma=0 \qquad \Longleftrightarrow \qquad  \frac{\partial\gamma^i}{\partial q^j}=\theta_j\gamma_i.   
\end{equation}
The next step is to compare $\wdt X_H$ with $T\gamma(\wdt X^\gamma_H)$. For $T\gamma(\wdt X^\gamma_H)$ we have
$$\tilde X^\gamma_H= \frac{\partial}{\partial t}+\frac{\partial H}{\partial p_i}\frac{\partial}{\partial q^i} $$
and
$$ T\gamma\Big(\frac{\partial}{\partial t}\Big)= \frac{\partial}{\partial t}+\frac{\partial\gamma^j}{\partial t}\frac{\partial}{\partial p_j}, \qquad     T\gamma\Big(\frac{\partial}{\partial q^i}\Big)=\frac{\partial}{\partial q^i}+\frac{\partial\gamma^j}{\partial q^i}\frac{\partial}{\partial p_j}  $$
so that 
$$ T\gamma(\wdt X^\gamma_H) =  \frac{\partial}{\partial t}+
\frac{\partial H}{\partial p_i} \frac{\partial}{\partial q^i}+\Big( \frac{\partial\gamma^j}{\partial t}+\frac{\partial H}{\partial p_i}\frac{\partial\gamma^j}{\partial q^i} \Big) \frac{\partial}{\partial p_j}.$$
Therefore, the condition $T\gamma(\wdt X^\gamma_H)=\wdt X_H$ in coordinates reads
\begin{equation}\label{equationHJ}
 -\frac{\partial H}{\partial q^i}+\frac{\partial H}{\partial p_j}(\theta_j\gamma_i-\theta_i\gamma_j)+\theta_iH    =\Big( \frac{\partial\gamma^i}{\partial t}+\frac{\partial H}{\partial p_j}\frac{\partial\gamma^i}{\partial q^j} \Big) .
\end{equation}
On the other hand, from ii) we have $[d_\theta(H\circ\gamma_t)]^V=-\dot\gamma_q^V$. In coordinates one has
$$  [d_\theta(H\circ\gamma_t)]^V= -\Big(\frac{\partial H}{\partial q^i}+\frac{\partial H}{\partial p_j}\frac{\partial\gamma^j}{\partial q^i}-\theta_i H\Big)\frac{\partial }{\partial p_i}, \qquad \quad  \dot\gamma_q^V= -\frac{\partial\gamma^i }{\partial t} \frac{\partial }{\partial p_i}  $$
so that 
\begin{equation}\label{equationgamma}
\frac{\partial\gamma^i }{\partial t} = -\frac{\partial H}{\partial q^i}-\frac{\partial H}{\partial p_j}\frac{\partial\gamma^j}{\partial q^i}+\theta_i H. 
\end{equation}
Now, it is a matter of straightforward computation to see that (\ref{equationHJ}) and (\ref{equationgamma}) are equivalent if and only if (\ref{dgammazero}) is satisfied.    
$$\qquad\qquad\qquad\qquad\qquad\qquad\qquad\qquad\qquad\qquad\qquad\qquad\qquad\qquad\qquad\qquad\qquad\qquad \blacksquare$$

Let us notice that the condition $[d_\theta(H\circ\gamma_t)+\dot\gamma_q]^V=0$ holds if and only if $d_\theta(H\circ\gamma_t)+\dot\gamma_q=0$. In coordinates we have
\begin{equation}\label{lcsHJequation}
\frac{\partial\gamma^i }{\partial t} + \frac{\partial H}{\partial q^i}+\frac{\partial H}{\partial p_j}\frac{\partial\gamma^j}{\partial q^i}-\theta_i H=0.
\end{equation}
We will refer to (\ref{lcsHJequation}) as the Hamilton-Jacobi equation on a locally conformal symplectic manifold.

\section{Applications: Contact structures and lcs HJ}

In order to shed some light on the applications of time-dependent locally conformal symplectic structures, we introduce the concept of contact pairs in the frame of contact geometry introduced in \cite{Bande,Bande2}. 

 A contact pair of type 
$(h,k)$ on a $(2h+2k+2)$-dimensional manifold is a pair of one-forms $(\alpha, \beta)$, such that 
$\alpha\land(d\alpha)^{h}\land\beta\land (d\beta)^{k}$ is a volume form, $(d\alpha)^{h+1}=0$ and $(d\beta)^{k+1}=0$. To this pair there are two associated Reeb vector fields $A$ and $B$, uniquely determined by the following  conditions: $\alpha(A)=\beta(B)=1$,
$\alpha(B)=\beta(A)=0$ and $i_{A}d\alpha=i_{A}d\beta=i_{B}d\alpha=i_{B}d\beta=0$. It turns out that a contact pair $(\alpha, \beta)$ of type $(h,0)$ gives rise to the lcs form 
\begin{equation}\label{cplcs}
\Omega=d\alpha+\alpha \land \beta .
\end{equation}
\noindent
Let us provide now a necessary and sufficient condition for a lcs form to arise from a contact pair. Considering contact pairs of type $(h,0)$, so that $\beta$ is a closed one-form, and the dimension of the manifold is $2h+2$, more generally, we can provide pairs of one-forms $(\alpha,\beta)$ such that $d\beta=0$ and $\alpha\land (d\alpha)^{h}\land\beta$ is a volume form. We shall refer to these pairs as {generalized contact pairs} of type $(h,0)$. For a generalized contact pair one defines a Reeb distribution $\mathcal{R}$ consisting of tangent vectors $Y$ satisfying the equation $(i_Yd\alpha )\vert_{\ker(\alpha)\cap\ker(\beta)}=0$.
    
Suppose that $\Omega$ is a lcs form, and $X$ is a vector field satisfying ${\mathcal L}_X\Omega =0$. Then, we have $ d\mathcal{L}_X\Omega= 0$, which by the nondegeneracy of $\Omega$ implies ${\mathcal L}_X\theta = 0$. It means, that $d (\theta (X))=0$, since $\theta$ is a closed form. Thus $\theta (X)=1$ is a constant that we have normalized. The following result is discussed in detail in \cite{vaisman}.

\noindent
\begin{theorem}
Let $M$ be a smooth manifold of dimension $2h+2$. There is a bijection between contact pairs $(\alpha,\beta)$ of type $(h,0)$ and lcs forms $\Omega$ with Lee form $\theta$, that admit a vector field $X$ satisfying ${\mathcal L}_X\Omega=0$ and $\theta (X)=1$. The bijection maps the Reeb vector fields $A$ and $B$ of $(\alpha,\beta)$ to $L$ and $X$ respectively. Moreover,
$\Omega^{h+1}$ and $\alpha\wedge (d\alpha)^h\wedge\beta$ define the same orientation on $M$.
\end{theorem}
\noindent
There is a partial generalization of this result to the case of generalized contact pairs in place of contact pairs \cite{Bande,Bande2}.

It turns out, that on a closed manifold, every generalized contact pair $(\alpha,\beta)$ of type $(h,0)$ gives rise to a lcs form
\begin{equation}\label{contactlcs}
\Omega = d\alpha +c\alpha\wedge\beta
\end{equation}
when $c\in\mathbb{R}$ is large. The Lee form $\theta$ of $\Omega$ equals $c\beta$. It is straightforward to check that $d\Omega=\Omega\wedge c\beta$. To check nondegeneracy we compute
$$
\Omega^{h+1} = (d\alpha +c\alpha\wedge\beta)^{h+1} = (d\alpha)^{h+1}+c(h+1)\alpha\wedge (d\alpha)^h\wedge\beta \ .
$$
For a large $c$, the second summand dominates the first summand,
so the right hand side is a volume form. 
In this case the equality $A=L$ no longer holds, in fact $L$ in general is not proportional to the Reeb vector field $A$.
The other Reeb vector field $B$ does not give an infinitesimal automorphism $X$ of the lcs form. If a closed manifold $M$ admits a (possibly generalized) contact pair $(\alpha,\beta)$, then the existence of the closed non-vanishing 
one-form $\beta$ implies that $M$ fibers over the circle.

We would like to show now some physical examples related to the existence of a lcs structure associated with contact pairs. Nilpotent Lie groups provide interesting examples of contact pairs \cite{7}, from which one can build a locally conformal symplectic two-form, as it is shown in the examples below. In order to describe the Lie algebra of a Lie group, we give only the non-zero ordered brackets of the fundamental fields $X_i$. The dual forms of $X_i$ will be denoted by $\eta^i$.

\subsection{Examples: Coupled linear differential equations}

Consider the  indecomposable nilpotent Lie algebra $\mathfrak{g}_{4,1}$ with nontrivial commutators
\begin{equation}\label{g41}
[X_1,X_4]=X_3,\qquad [X_1,X_3]=X_2.
\end{equation}
The pair $(\eta^2,\eta^4)$ determines a contact pair of type $(1,0)$ on the corresponding Lie group, where $\eta^i$ are the dual forms of the vector fields $X_i$ in \eqref{g41}. One can construct a locally conformally symplectic form that is
\begin{equation}\label{constructionlcs}
 \Omega=d\eta^2+\eta^2\wedge \eta^4.
 \end{equation}
and the Lee form is $\theta=\eta^4$.
The Lie algebra \eqref{g41} admits different representations depending on the dimension of the space we are working on \cite{PBNL}. Let us select one that is convenient for the time-dependent lcs case.

Using the representations ${\bf 1}$, ${\bf 2}$ and ${\bf 4}$ given in \cite{PBNL} completing the basis for the Lie algebra $\mathfrak{g}_{4,1}$, we can build up linear systems of differential equations in the five-dimensional manifold coordinated by $(t,x_1,x_2,x_3,x_4)$.

\subsubsection*{System 1}
 Using representation ${\bf 1}$, given by
\begin{equation}\label{rep1}
\{X_1=\partial_{x_2},X_2=\partial_{x_1}, X_3=x_2\partial_{x_1}+x_3\partial_{x_2}+\partial_{x_4},X_4=\partial_{x_3}\} 
\end{equation}
we obtain a time-dependent vector field
$$X_t=a_1(t)\partial_{x_1}+a_2(t)\partial_{x_2}+a_3(t)\partial_{x_3}+a_4(t)\left(x_2\partial_{x_1}+x_3\partial_{x_2}+\partial_{x_4}\right),$$ which can be rewritten as
\begin{equation}
X_t=\left(a_1(t)+a_4(t)x_2\right)\partial_{x_1}+\left(a_2(t)+a_4(t)x_3\right)\partial_{x_2}+a_3(t)\partial_{x_3}+a_4(t)\partial_{x_4}
\end{equation}
and whose integral curves correspond with the system
\begin{equation}\label{sysrep1}
\begin{cases}
\quad \dot{x}_1=a_1(t)+a_4(t)x_2,\nonumber \\
\quad \dot{x}_2=a_2(t)+a_4(t)x_3,\nonumber\\
\quad \dot{x}_3=a_3(t),\nonumber\\
\quad \dot{x}_4=a_4(t).
\end{cases}
\end{equation}
The dual basis of one-forms reads
\begin{equation}\label{dualrep1}
 \{\eta^1=dx_2-x_3dx_4,\eta^2=dx_1-x_2dx_4,\eta^3=dx_4,\eta^4=dx_3\}
\end{equation}
\noindent
Using above forms we can construct a locally conformal symplectic two-form as in \eqref{constructionlcs}. In this case, it reads
\begin{equation}\label{w1lcs}
\Omega_{{\bf 1}}=-dx_2\wedge dx_4+dx_1\wedge dx_3-x_2dx_4\wedge dx_3.
\end{equation}
It is straightfoward to see that \eqref{w1lcs} fulfills $d\Omega_{{\bf 1}}=\eta^4\wedge\Omega_{{\bf 1}} $. Therefore, it is a lcs form with a Lee form  $\theta=\eta^4$.

\subsubsection*{System 2}

Using representation ${\bf 2}$, which reads 
\begin{equation}\label{rep2}
\{X_1=\partial_{x_2}, X_2=\partial_{x_1},X_3=x_2\partial_{x_1}+x_3\partial_{x_2}+x_4\partial_{x_3}+\partial_{x_4}, X_4=\partial_{x_3}\}
\end{equation}
we obtain a time-dependent vector field 
$$X_t=a_1(t)\partial_{x_1}+a_2(t)\partial_{x_2}+a_3(t)\partial_{x_3}+a_4(t)\left(x_2\partial_{x_1}+x_3\partial_{x_2}+x_4\partial_{x_3}+\partial_{x_4}\right),$$
which can be rewritten as
\begin{equation}
X_t=\left(a_1(t)+a_4(t)x_2\right)\partial_{x_1}+\left(a_2(t)+a_4(t)x_3\right)\partial_{x_2}+(a_3(t)+a_4(t)x_4)\partial_{x_3}+a_4(t)\partial_{x_4}
\end{equation}
and whose integral curves correspond with the system
\begin{equation}\label{sysrep2}
\begin{cases}
\quad \dot{x}_1=a_1(t)+a_4(t)x_2,\nonumber\\
\quad \dot{x}_2=a_2(t)+a_4(t)x_3,\nonumber\\
\quad \dot{x}_3=a_3(t)+a_4(t)x_4,\nonumber\\
\quad \dot{x}_4=a_4(t).
\end{cases}
\end{equation}
The dual basis of one-forms is
\begin{equation}
\{\eta^1=dx_2-x_3dx_4, \eta^2=dx_1-x_2dx_4, \eta^3=dx_4,\eta^4=dx_3-x_4dx_4 \}
\end{equation}
Again, we can construct a locally conformal symplectic two-form according to \eqref{constructionlcs}
\begin{equation}\label{w4lcs}
\Omega_{{\bf 2}}=-dx_2\wedge dx_4+dx_1\wedge dx_3-x_4dx_1\wedge dx_4+x_2dx_3\wedge dx_4
\end{equation}
with a Lee form $\theta=\eta^4$. It is easy to check that $d\Omega_{{\bf 2}}=\theta\wedge\Omega_{{\bf 2}}$.

\subsubsection*{System 4}

\noindent
We use representation ${\bf 4}$, which reads 
\begin{equation}\label{rep4}
\{X_1=\partial_{x_2},X_2=\partial_{x_1},X_3=x_2\partial_{x_1}+x_4\partial_{x_3}-\partial_{x_4}, X_4=x_3\partial_{x_1}+x_4\partial_{x_2}+\partial_{x_3} \}.
\end{equation}
\noindent
Again, we obtain a $t$-dependent vector field $$X_t=a_1(t)\partial_{x_1}+a_2(t)\partial_{x_2}+a_3(t)\left(x_3\partial_{x_1}+x_4\partial_{x_2}+\partial_{x_3}\right)+a_4(t)\left(x_2\partial_{x_1}+x_4\partial_{x_3}-\partial_{x_4}\right)$$
which can be rewritten as
\begin{equation*}
X_t=\left(a_1(t)+a_3(t)x_3+a_4(t)x_2\right)\partial_{x_1}+\left(a_2(t)+a_3(t)x_4\right)\partial_{x_2}+(a_3(t)+a_4(t)x_4)\partial_{x_3}-a_4(t)\partial_{x_4}
\end{equation*}
and whose integral curves correspond with the system
\begin{equation}\label{sysrep3}
\begin{cases}
\quad \dot{x}_1=a_1(t)+a_3(t)x_3+a_4(t)x_2,\nonumber\\
\quad \dot{x}_2=a_2(t)+a_3(t)x_4,\nonumber\\
\quad \dot{x}_3=a_3(t)+a_4(t)x_4,\nonumber\\
\quad \dot{x}_4=-a_4(t).
\end{cases}
\end{equation}
In this case, the basis of dual one-forms is
\begin{equation*}
\{\eta^1=dx_2-x_4dx_3-x_4^2dx_4,\eta^2=dx_1-x_3dx_3-(x_3x_4-x_2)dx_4,\eta^3=-dx_4,\eta^4=dx_3+x_4dx_4\}
\end{equation*}
and the associated lcs two-form $\Omega=d\eta^2+\eta^2\wedge\eta^4$ reads
\begin{equation}\label{w4lcs}
\Omega_{{\bf 4}}=dx_1\wedge dx_3+x_4dx_1\wedge dx_4+dx_2\wedge dx_4-(x_4+x_2)dx_3\wedge dx_4
\end{equation}
where the Lee form is $\theta=\eta^4$.

These systems are related to control systems on $\mathbb{R}^5\times \mathbb{R}$ when one reduces the system by the symmetry coordinate $x_5$ in \cite{LS}. For example, in {\bf System 1}, the functions $a_3(t)$ and $a_4(t)$ are the controls allowing the dynamics to connect two points in $\mathbb{R}^4$ while optimazing a cost functional depending on the controls. These systems, under certain particular considerations, are equivalent to plate-ball systems with respect to the optimization problem \cite{Brockett}.

\subsection{HJ equation for a system with a time-dependent potential}


Consider a planar Euclidean space $Q=\mathbb R^n$. In order to define a lcs manifold structure on the cotangent bundle, we assume that we have a closed one-form $\theta=\theta_idq^i$, which is the pull-back of the one-form $\vartheta$ on $Q$. Then $T_\theta^*Q\times\mathbb R=\mathbb R^{2n}\times\mathbb R$ and
\begin{equation} \label{Ex-Omega}
\Omega_\theta=\Omega_Q+\theta\wedge \theta_Q=dq^i\wedge dp_i - \theta_ip_jdq^i\wedge dq^j.
\end{equation}
Let us consider a system described by a Hamiltonian with a time-dependent potential 
\begin{equation} \label{Ex-ham-func}
H=\frac{1}{2m}p_ip^i+V(q,t).
\end{equation}
Applying the time-dependent lcs HJ equation given in \eqref{tlcsHJ}, we obtain a solution $\gamma(t,q^i)=(t,q^i,\gamma^k(t,q^i))$ for the Hamilton equations \eqref{tHameq} associated with \eqref{Ex-ham-func}. From (\ref{lcsHJequation}) we have that lcs HJ equation for $\gamma$ reads
$$\frac{\partial\gamma^i }{\partial t}+\frac{1}{2m}p_j\frac{\partial\gamma^j}{\partial q^i} + \frac{\partial V}{\partial q^i}-\theta_i\Big(\frac{1}{2m}p_ip^i+V(q,t)\Big)=0.$$


\section{Acknowledgements}


The research of Marcin Zając was financially supported by the University’s Integrated Development Programme (ZIP) of University of Warsaw co-financed by the European Social Fund under the Operational Programme Knowledge Education Development 2014 - 2020, action 3.5.

\end{document}